\documentclass[aps,prc,preprint,groupedaddress]{revtex4-1}

\usepackage{graphics}
\usepackage{multirow}

\begin{document}


\title{Determination of the Quark Content of Scalar Mesons Using Hydrodynamical Flow in Heavy Ion Collisions}


\author{M.~Wussow}
\affiliation{Department of Physics, Augustana College, Sioux Falls, SD 57197}
\author{N.~Grau}
\email[]{ngrau@augie.edu}
\affiliation{Department of Physics, Augustana College, Sioux Falls, SD 57197}


\date{\today}

\begin{abstract}
We study the possibility of determining the quark content of the scalar mesons $a^0(980)$ and $f^0(980)$ through their hydrodynamical flow signature in relativistic heavy ion collisions. Utilizing the constituent quark scaling of hydrodynamic flow, we find that the tetraquark $a^0(980)$ or $f^0(980)$ mesons will have a $v_2$ of 0.38 at transverse momentum of 6 GeV/c in 20-60\% central Au+Au collisions at $\sqrt{s_{NN}}$ = 200 GeV. The feasibility of measuring $a^0(980)\rightarrow\pi^0\eta$ and $f^0(980)\rightarrow\pi^+\pi^-$ into the PHENIX and STAR detectors at the Relativistic Heavy Ion Collider (RHIC) is also discussed. Even though the mid-rapidity cross sections for these mesons at high-$p_T$ are non-negligible, their broad mass range will make them difficult to detect in both $p+p$ and Au+Au collisions.
\end{abstract}


\maketitle

\section{Introduction}\label{sec:introduction}
Nearly all hadrons that have been measured can be described as a color singlet state of either three quarks or anti-quarks or a quark-anti-quark pair. However, the theory of Quantum Chromodynamics (QCD) allows more exotic structures to exist, like tetraquarks ($q\bar{q}q\bar{q}$) \cite{Jaffe:1976ig}, pentaquarks ($qqqq\bar{q}$) \cite{Diakonov:1997mm} or multi-gluon states (glueballs) \cite{Morningstar:1999rf}. Several unexpected charm states have been discovered recently. For example, the $X$(3872) \cite{Choi:2003ue} and $Z^+$(4330) \cite{Choi:2007wga}, have not satisfactorily been explained as two-quark states but could be tetraquark states \cite{Ebert:2005nc,Rosner:2007mu}.

Not only could exotic quark states exist at the higher masses, but several well-established low mass resonances are candidates for multiquark states. In Ref.~\cite{Jaffe:1976ig}, Jaffe showed that the lowest mass scalar $q\bar{q}q\bar{q}$ states form a nonet, several of which describe established particles. High-statistics measurements from radiative $\phi$ decays to the $f^0$(980)~\cite{Aloisio:2002bt} and $a^0$(980)~\cite{Ambrosino:2009py} indicate that these are consistent with four-quark states.

Assigning valence quark content for a particle is based on the determination of its quantum numbers. It would be extremely useful to have a more direct measurement of their valence quark structure. It appears that data from relativistic heavy ion collisions from the Relativistic Heavy Ion Collider (RHIC) at Brookhaven National Laboratory have yielded a dynamical observable that is sensitive to the number of valence quarks of a hadron.

\begin{figure}[t!]
\scalebox{0.40}{\includegraphics{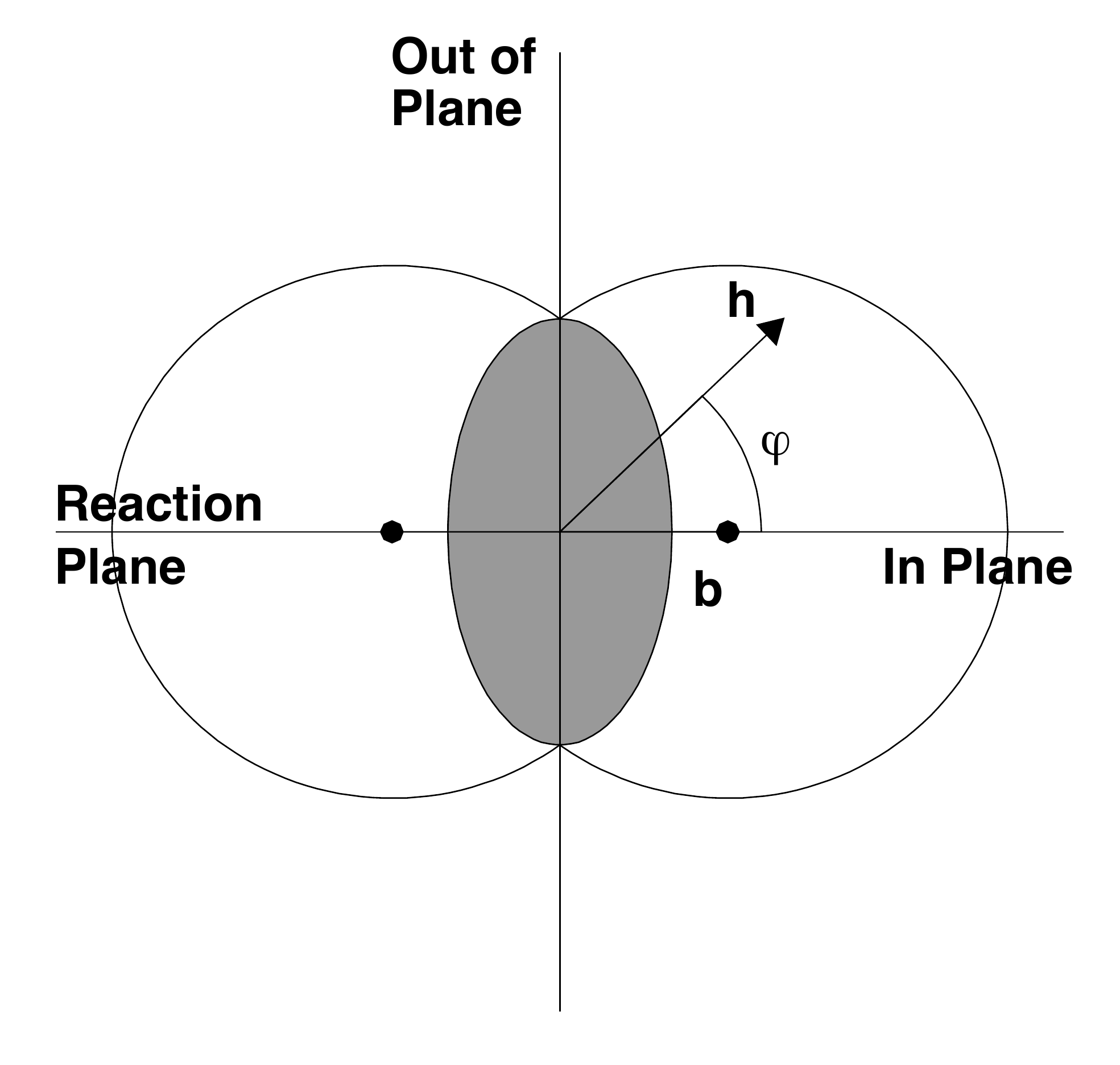}}
\caption{A beam view of a relativistic heavy ion collision. The impact parameter and the beam axis (into the page) form the reaction plane. The shaded, almond-shaped overlap region is the source of particle production, which is asymmetric in $\varphi$.}\label{fig:rp}
\end{figure}

In collisions of gold nuclei at $\sqrt{s_{NN}} =$ 200 GeV, a quark-gluon plasma is formed and behaves as a strongly-interacting perfect fluid~\cite{Adcox:2004mh,Adams:2005dq}. In collisions with a non-zero impact parameter, the overlap region forms an almond-like shape. This is depicted in Fig.~\ref{fig:rp}. Because of thermodynamic pressure gradients in this region, particle production is not azimuthally symmetric. More particles are emitted along the reaction plane, the plane resulting from the beam and impact parameter axes, than perpendicular to it. This azimuthal asymmetry is characterized by the coefficients of a Fourier expansion of the invariant yield~\cite{Voloshin:1994mz}
\begin{eqnarray}\label{eq:fourier}
\frac{1}{p_T}\frac{d^3N}{dp_Td\varphi dy} & = & \frac{1}{2\pi p_T}\frac{d^2N}{dp_Tdy}\nonumber \\
& & \times \left[1+2\sum_{n}v_{n}(p_T,y)\cos \left(n\varphi\right)\right],
\end{eqnarray}
where $p_T$, $y$, and $\varphi$ are the particle's momentum transverse to the beam direction, rapidiy, and angle with respect to the reaction plane, respectively. The $v_n(p_T,y)$ encode the strength of the azimuthal asymmetry.

When fluctuations in the initial energy and entropy density of the overlap region are small, symmetry requires only even Fourier coefficients to be non-zero. The dominant term in the series is $v_2$. Both the PHENIX and STAR experiments have measured a large number of identified baryon and meson $v_2$ as a function of $p_T$~\cite{Afanasiev:2007tv,Adare:2010sp,Abelev:2007rw,Abelev:2008ed}. These data follow a universal trend in $v_2/n_q$ vs. $KE_T/n_q$ (see Fig.~\ref{fig:phenixv2} and \ref{fig:starv2}). Here, $n_q$ is the number of valence quarks of the particle and $KE_T$ = $m_T-m$ = $\sqrt{p_T^2+m^2}-m$ is the transverse kinetic energy of the particle. Since this data fits light-flavor (up, down, and strange) baryons and mesons for $KE_T/n_q \lesssim$ 1 GeV, a light-flavored particle's $v_2$ is sensitive to the number of valence quarks of that particle.

Such a result can be explained quantitatively by the recombination model of hadronization~\cite{Fries:2003vb}. When the quark-gluon plasma cools, hadronization must take place. The recombination model allows for quarks nearby in phase space to cluster and, with appropriate quantum numbers, form hadrons. The first model of recombination explained the measured enhancement of the $p/\pi^+$ ratio in Au+Au collisions compared to p+p collisions. Because of the steeply falling momentum spectrum of quarks, the ability for three quarks of approximately equal momentum combining to form a proton is more favorable than two combining to form a meson with the same final momentum as the proton.

In such a model, the quark $v_2$ will become the hadron $v_2$ based on the combination of the quarks that take place. The initial suggested scaling of $v_2$ was~\cite{Molnar:2003ff}
\begin{equation}\label{eq:v2pocket}
v_2(p_T) \approx n_qv_2^{q}(p_T/n_q)
\end{equation}
However, detailed measurements of $v_2$ showed this scaling was incomplete and the $KE_T$ scaling was found to work better experimentally~\cite{Abelev:2007rw}. A more detailed recombination model, where one considers resonance-like scattering in the pre-hadronic state, which conserves energy and momentum in detail, is able to reproduce the observed $KE_T$ scaling~\cite{Ravagli:2007xx,Ravagli:2008rt,He:2010vw}.

With the discovery of constituent quark scaling of hadron $v_2$ at RHIC, we would like to search for evidence of particles with exotic quark structures. In this paper we present the study of the flow of light scalar mesons, $f_0(980)$ and $a_0(980)$, at RHIC. If the recombination model of hadronization is responsible for the translation of $v_2$ of quarks to $v_2$ of hadrons at RHIC, then these mesons will follow the observed scaling of $v_2$ and their number of valence quarks will then be known. A definitive measure of a tetraquark state would be an important result for QCD.

\section{Flow of Scalar Mesons}\label{sec:Flow}

\begin{figure}
\scalebox{0.40}{\includegraphics{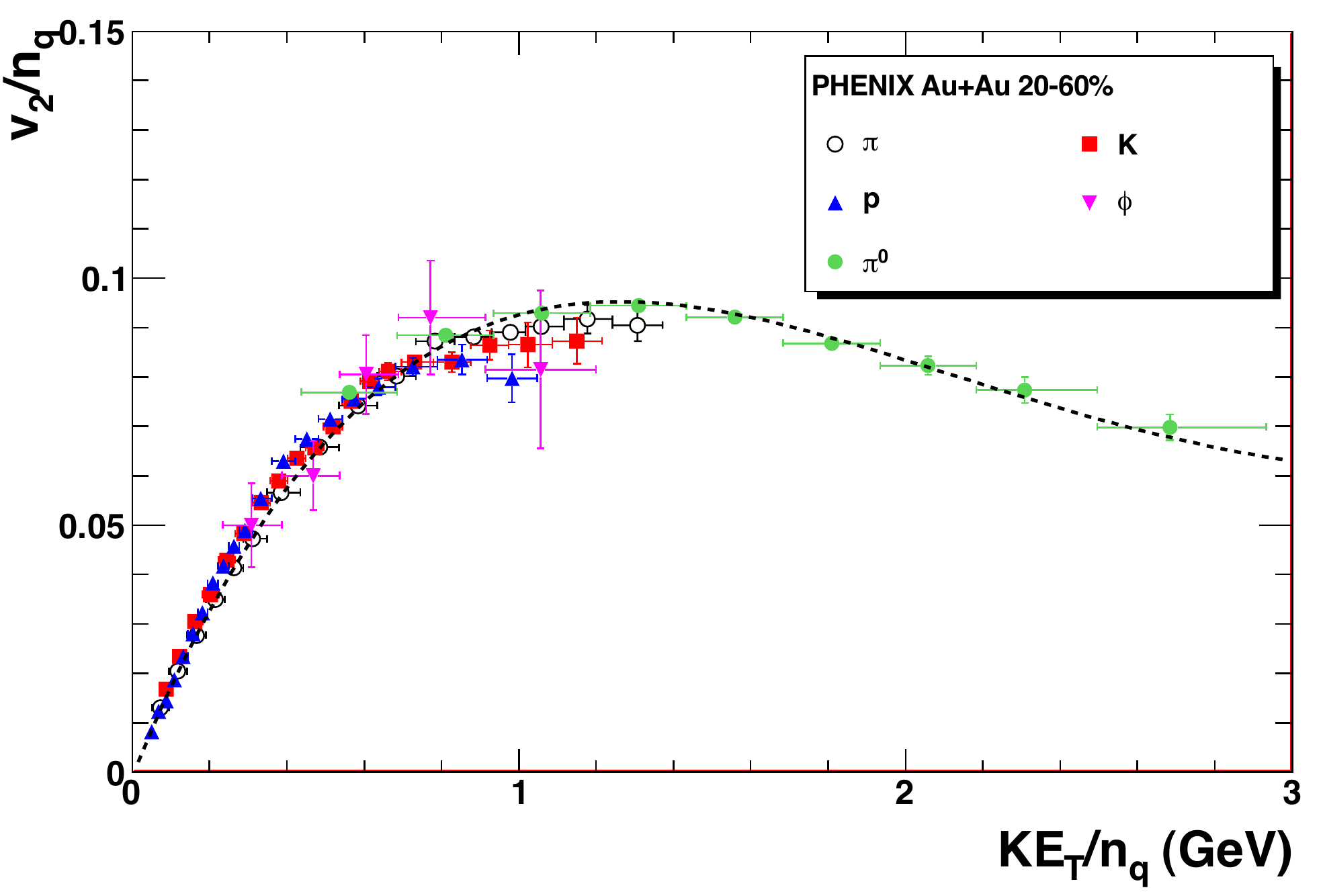}}
\caption{(Color online) The combined PHENIX $v_2$ data~\cite{Afanasiev:2007tv,Adare:2010sp} for identified $\pi^{\pm}$ (open circles), $\pi^0$ (green closed circles), $K^{\pm}$ (red squares), $p(\bar{p})$ (blue triangles), and $\phi$ (magenta inverted triangles) in 20-60\% central Au+Au collisions at $\sqrt{s_{NN}}$ = 200 GeV. The black line is a fit to the combined pion data.}
\label{fig:phenixv2}
\end{figure}

\begin{figure}
\scalebox{0.40}{\includegraphics{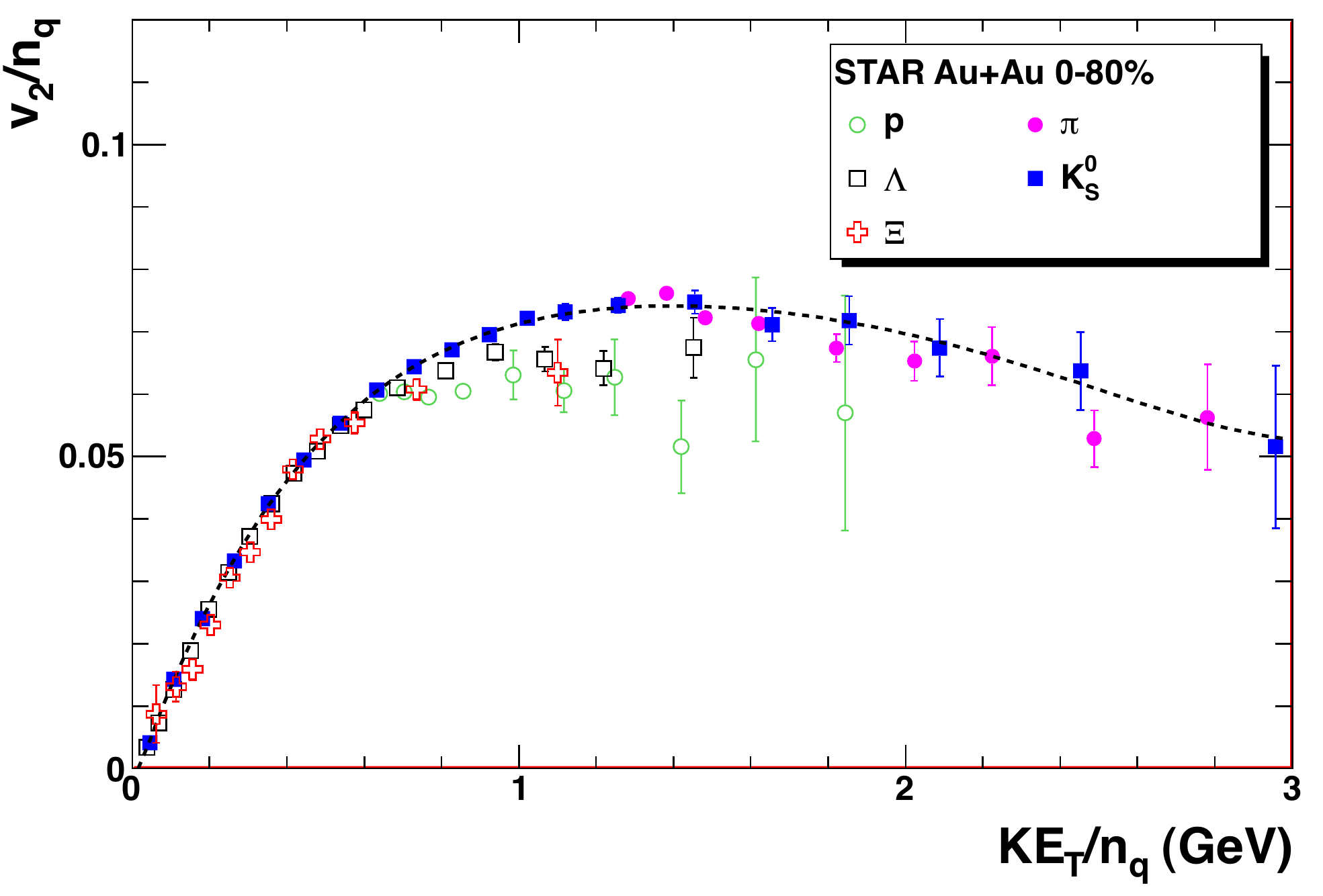}}
\caption{(Color online) The combined STAR $v_2$ data~\cite{Abelev:2008ed} for identified protons (green open circles), $\Lambda$ (open squares), $\Xi$ (red crosses), $\pi^{\pm}$ (magenta circles), and $K^0_S$ (blue squares) in 0-80\% Au+Au collisions at $\sqrt{s_{NN}}$ = 200 GeV. The dashed line is a fit to the kaon data.}
\label{fig:starv2}
\end{figure}

Using data it is possible to determine an expectation for the $v_2$ signature of the $a^0(980)$ and $f^0(980)$. Fig.~\ref{fig:phenixv2} shows the compilation of PHENIX data for $v_2$ of identified hadrons in 20-60\% central Au+Au collisions at $\sqrt{s_{NN}}$ = 200 GeV~\cite{Afanasiev:2007tv,Adare:2010sp}. The $v_2/n_q$ is plotted as a function of $KE_{T} = m_T-m$. Up to $KE_T/n_q$ of approximately 1 GeV, all of the data follow the same trend. The PHENIX $\pi^0$ data were published in 10\% centrality bins. They were combined for the figure using an eccentricity weighted average of the appropriate centrality bins. The eccentricity was determined by a Glauber Monte Carlo of the Au+Au collisions~\cite{Miller:2007ri}. We fit a 4$^{\rm th}$-order polynomial to the combined pion data: $\pi^{\pm}$ for $KE_T/n_q<$ 1 GeV and $\pi^0$ for $KE_T/n_q>$ 1 GeV. This is shown as the dashed line in Fig.~\ref{fig:phenixv2}. Fig.~\ref{fig:starv2} is a compilation of identified hadron $v_2$ from STAR in minimum bias (0-80\%) central Au+Au collisions at $\sqrt{s_{NN}}$ = 200 GeV~\cite{Abelev:2008ed}. A fit to the $K^0_S$ data, which exists over the broadest range in $KE_T/n_q$, is also shown as the dashed line.

\begin{figure}
\scalebox{0.4}{\includegraphics{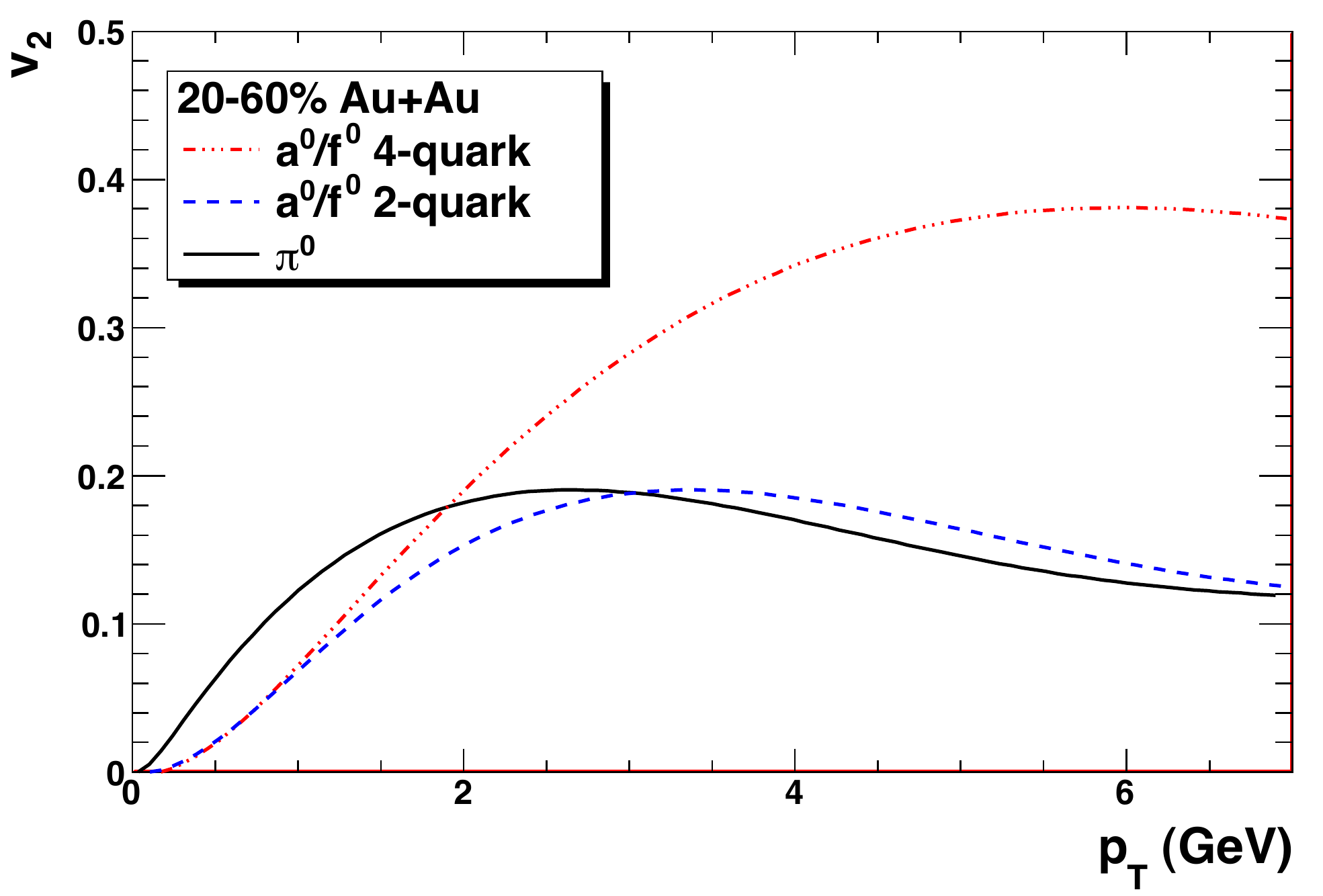}}
\caption{(Color online) The extrapolated $v_2$ of the $a^0(980)$ or $f^0(980)$ in 20-60\% Au+Au collisions as a function of transverse momentum if they are either 2-quark states (dashed blue line) or 4-quark states (red dot-dashed line). For comparison the PHENIX pion $v_2$ is given as the solid (black) line.}
\label{fig:a0f0v2}
\end{figure}

\begin{figure}
\scalebox{0.40}{\includegraphics{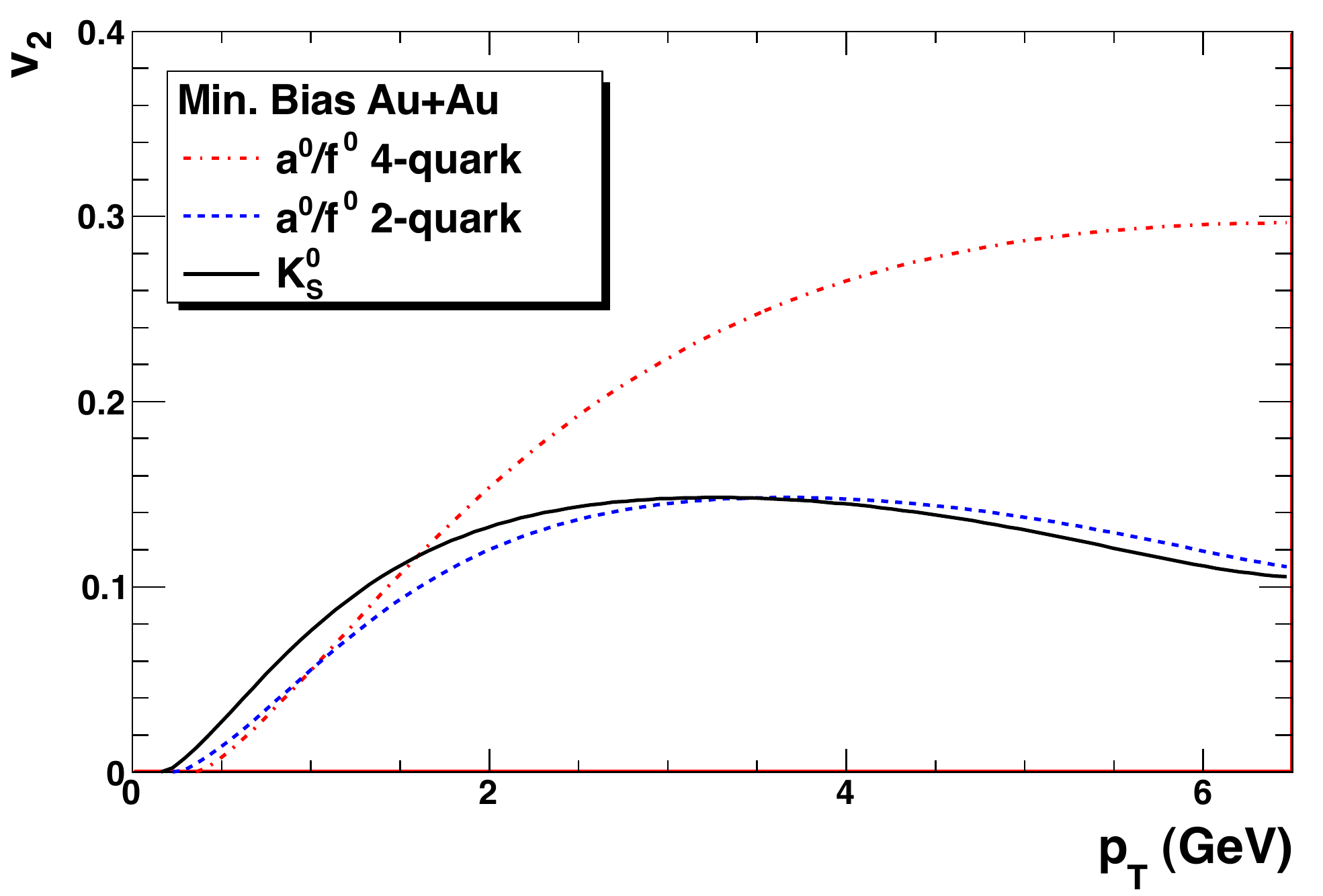}}
\caption{(Color online) The extrapolated $v_2$ of the $a^0(980)$ or $f^0(980)$ in minimum bias (0-80\%) Au+Au collisions as a function of transverse momentum if they are either 2-quark states (dashed blue line) or 4-quark states (red dot-dashed line). For comparison the STAR $K^0_S$ $v_2$ is given as the solid (black) line.}
\label{fig:a0f0v2_minbias}
\end{figure}

From here it is straight-forward to undo the scaling and plot the $v_2$ as a function of $p_T$ by knowing the mass and the number of valence quarks of the particle of interest. This is shown in Fig.~\ref{fig:a0f0v2} for 20-60\% collisions and Fig.~\ref{fig:a0f0v2_minbias} for 0-80\% collisions for a particle of mass 980 MeV and consisting of either two (dashed line) or four quarks (dot-dashed line). There is a very large separation between these two curves. It is rougly consistent with the pocket formula Eq.~\ref{eq:v2pocket}, where the 4-quark $v_2$ curve is roughly twice the 2-quark $v_2$, especially at intermediate $p_T$ where the recombination model dominates the particle production. In the mid-central collisions (20-60\%), the 4-quark $v_2$ at 6 GeV is 2.6 times larger than the 2-quark $v_2$ and reaches a value of 0.38, a $v_2$ that dwarfs all other $v_2$ measurements from light hadrons. Even in minimum bias collisions the 4-quark $v_2$ reaches a value of 0.30 at 6 GeV and is 2.5 times higher than the 2-quark $v_2$.

We note that, in both the minimum bias and in mid-central collisions, the 4-quark $v_2$ peaks between $p_T$ of 5-6 GeV or $KE_T/n_q$ = 1-1.2 GeV. Therefore, the entire curve is within the measured $KE_T$ scaling of $v_2$ at RHIC. We expect the 2-quark $v_2$ to follow the trend of the $\pi^0$ or the $K^0_S$ and decrease as a function of $p_T$, widening the gap between the expected $v_2$ signatures.

\section{Feasibility of Measuring Scalar Mesons At RHIC}\label{sec:Yield}

In order to establish that a measurement of the $v_2$ can be made, we must get a sense of the signal strength. Important in $v_2$ measurements of resonances is the ability to measure the signal and the background $v_2$. In this section, we estimate the signal-to-background in $p+p$ and Au+Au collisions to determine the feasibility of the measurement. 

In Au+Au collisions there will be at least three production sources: from radiative $\phi$ decays, ``prompt'' production from fragmentation, and production from recombination. We take the approach of estimating the first two components of these production mechanisms in $p+p$ collisions. From this we can evaluate if the measurement in heavy ion collisions is feasible.

Specifically, we are interested in the cross section of these mesons at high $p_T$, \textit{i.e.}~at 6 GeV where there is maximal difference between the 2-quark and 4-quark $v_2$. We focus on the RHIC experiments where these mesons would be measured. The $a^0(980)\rightarrow\pi^0\eta$ decay would be possible in PHENIX where their calorimeter has measured both $\pi^0$~\cite{Adare:2008qa} and $\eta$~\cite{Adare:2010dc} inclusive production in $p+p$ and heavy ion collisions. The $f^0(980)\rightarrow\pi^+\pi^-$ would be a favorable decay for STAR, which has particle identification capabilities over a large kinematic range~\cite{Abelev:2006jr}.


To estimate the cross section, we first study the radiative $\phi$ decays. PHENIX has measured the cross section of $\phi$ and several other neutral mesons in p+p collisions at $\sqrt{s}$ = 200 GeV~\cite{Adare:2010fe}. The spectral shape for all mesons is well described with a Tsallis distribution~\cite{Tsallis:1987eu},
\begin{eqnarray}\label{eq:tsallis}
E\frac{d^3\sigma}{dp^3} & = & \frac{1}{2\pi}\frac{d\sigma}{dy}\frac{(n-1)(n-2)}{(nT+m_{0}(n-1))(nT+m_{0})} \nonumber \\
&  & \times (\frac{nT+m_T}{nT+m_{0}})^{-n}
\end{eqnarray}
where $d\sigma/dy$ is the integrated cross section of the particle production at midrapidity. We modeled the radiative $\phi\rightarrow a^0(980)\gamma$ decay using the measured spectral shape from PHENIX: $d\sigma/dy$ = 0.42 mb, $n$ = 10, and $T$ = 120 MeV. With the measured branching fraction from KLOE (7$\times10^{-5}$), the $d\sigma/dy$ for 6 GeV $a^0(980)$ production into the PHENIX pseudorapidity acceptance from $\phi$ decays is 9 nb. Requiring that the $\eta$ daughter decays to 2 photons, which is the most efficient method to measure $\eta$ mesons in PHENIX, reduces the cross section to 3 nb. If, instead, the $\phi$ would radiatively decay into an $f^0(980)$, based on the measured branching fraction from KLOE ($10^{-4}$), $d\sigma/dy$ for 6 GeV $f^0(980)$ into STAR would be 58 nb. If you require the $f^0(980)$ to decay into two pions, this reduces the cross section to 19 nb. These cross sections, therefore, represent the lower bounds on the production cross section for these scalar mesons into PHENIX and STAR, respectively.

There should be an additional prompt production of $a^0(980)$ and $f^0(980)$ mesons from fragmentation. This has not been measured at collider energies. We estimate the prompt cross section the following way. If the mesons are two-quark state, because their mass and quark content are similar to that of the $\eta^{\prime}$, their prompt production will be similar. PHENIX has measured the $\eta^{\prime}$ production cross section in p+p collisions~\cite{Adare:2010fe}. Its spectral shape is also given by the Tsallis distribution, Eq.~\ref{eq:tsallis}. We use the PHENIX parametrization, $d\sigma/dy$ = 0.7 mb, $n$ = 10, and $T$ = 120 MeV, which yields a $d\sigma/dy$ = 87 nb at 6 GeV. Requiring the $\eta$ from the $a^0(980)$ decay into two photons reduces the cross section to 34 nb. The $f^0(980)$ decaying into two pions reduces the cross section to 29 nb.

For a lower bound estimate of the prompt cross section, we use the following argument. If the mesons are 4-quark states, their production compared to 2-quark fragmentation would go approximately as the $(p/\pi^+)^2$. This is because this ratio encodes the difference in the string fragmenting into di-quark-anti-di-quark compared to quark-anti-quark pairs. In $p+p$ collisions, this ratio is approximately 0.1. So we would expect that a lower estimate of the prompt cross section would be a factor of 100 less than the 2-quark production. In this case, the cross section for 6 GeV $a^0(980)$ into PHENIX and 6 GeV $f^0(980)$ into STAR is 0.34 nb and 0.29 nb, respectively. It is interesting to note that because the $\eta^\prime$ cross section at 6 GeV is the same order of magnitude as the cross section from $\phi$ decays, the lower bound values are significantly less than the radiative $\phi$ production. Therefore, the $\phi$ decays would really represent an absolute lower bound for their production in $p+p$ collisions.

\begin{figure}
\scalebox{0.4}{\includegraphics{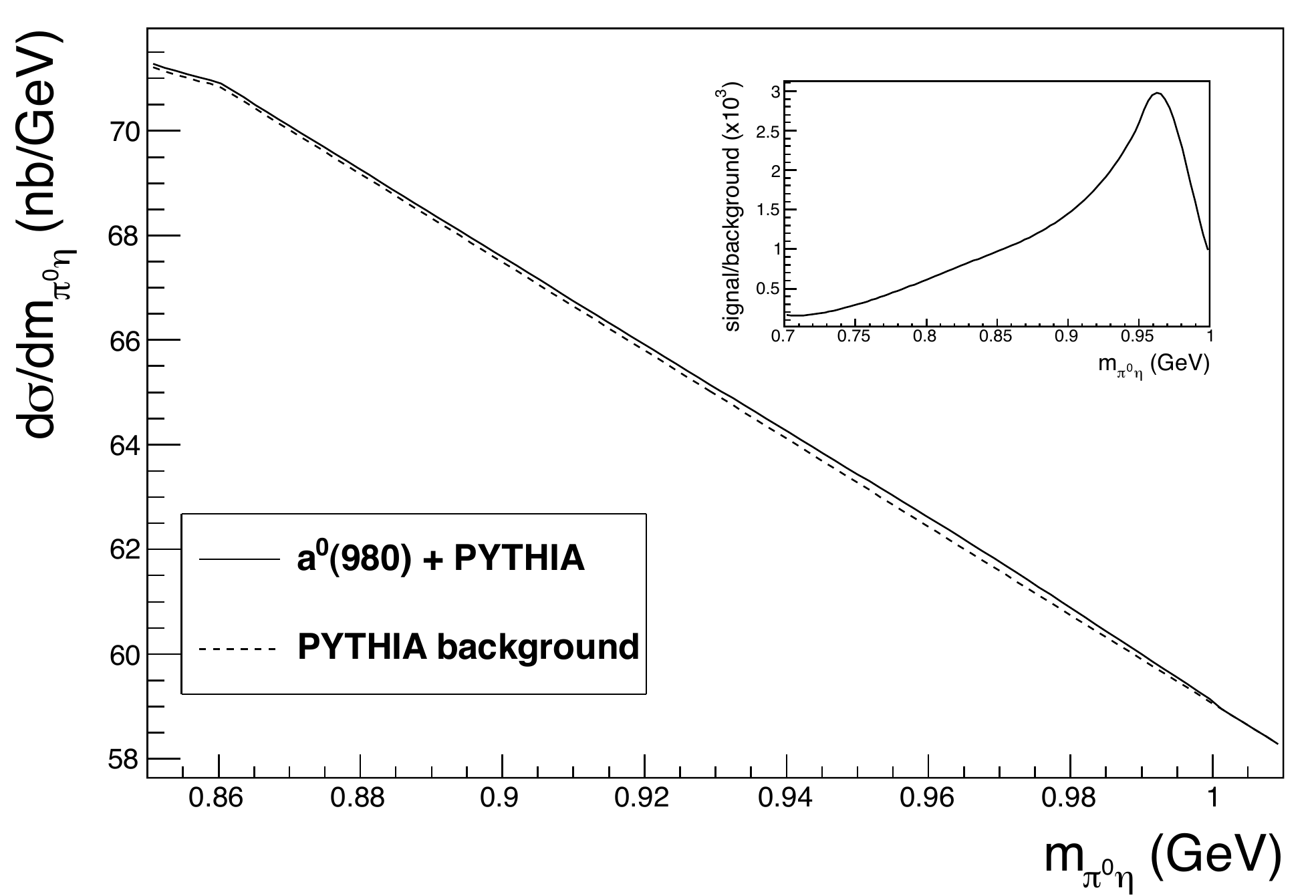}}
\caption{The 6 GeV pair $p_T$ $\pi^0\eta$ invariant mass distribution for background from PYTHIA (dashed) and the parametrized $a^0(980)$ signal (solid) assuming the same cross section as the $\eta^\prime$. In both the signal and the background, the $\pi^0$ and $\eta$ decay to two photons and are required to be within the PHENIX $y$ acceptance. The inset is the signal-to-background ratio over the full mass range of the $a^{0}(980)$.}
\label{fig:a0signalphenix}
\end{figure}

To find the signal-to-background ratio, we generated PYTHIA p+p events at $\sqrt{s}$ = 200 GeV with Tune A~\cite{Field:2002vt}.  For the $a^0(980)$ background all $\pi^0$-$\eta$ pairs, each decaying to two photons, were combined. Since this will be measured into the PHENIX acceptance, we also applied $p_T$ cuts consistent with the PHENIX capabilities ~\cite{Adare:2008qa,Adare:2010dc}. We required $\pi^0$ $p_T>$ 1 GeV, $\eta$ $p_T>$ 2 GeV, and $|y|<$ 0.4 for both particles in the signal and background. These cuts reduced the signal cross section to 22 nb. The signal peak was parametrized from the KLOE data and normalized to have this cross section. The combined signal and background mass distribution is shown in Fig.~\ref{fig:a0signalphenix}. If the $a^0(980)$ would have a cross section similar to the $\eta^\prime$, then a peak is barely visible in the invariant mass distribution of $\pi^0-\eta$ pairs. If one integrates over the full mass range of 0.7-1.0 GeV, the signal-to-background is $10^{-3}$. The inset in Fig.~\ref{fig:a0signalphenix} shows the signal-to-background as a function of invariant mass. It reaches a maximum of $3\times10^{-3}$ near 980 MeV. It follows that if the $\phi$ decay is the dominant production mechanism, the signal-to-background will be reduced. With the $p_T$ cuts above, the signal from $\phi$ decay reduces to 2 nb over the whole mass range and the signal-to-background would be reduced to 5$\times10^{-5}$. Even though the cross section is small but measurable with RHIC luminosities, the fact that the resonance is spread over 300 MeV will make the measurement difficult. Additional cuts would be required to improve the signal-to-background.

\begin{figure}
\scalebox{0.4}{\includegraphics{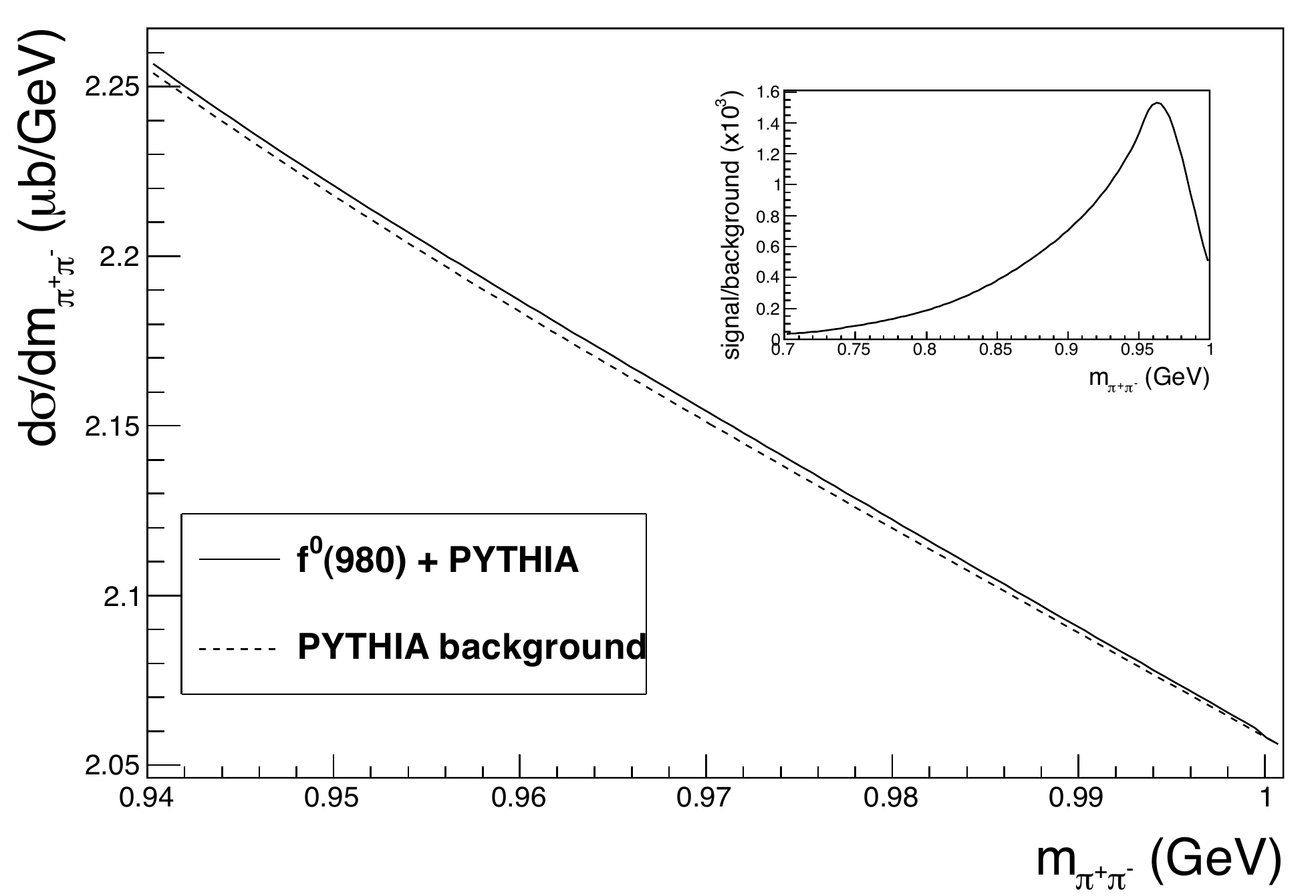}}
\caption{The 6 GeV pair $p_T$ $\pi^+\pi^-$ invariant mass distribution for background from PYTHIA (dashed) and the parametrized $f^0(980)$ signal (solid) assuming the same cross section as the $\eta^\prime$. In both the signal and the background, the $\pi^\pm$ are required to be within the STAR $y$ acceptance. The inset is the signal-to-background ratio over the full mass range of the $f^{0}(980)$.}
\label{fig:f0signalstar}
\end{figure}

With the same PYTHIA background, we studied the $f^0(980)$ signal-to-background into the STAR acceptance. We assume the following cuts on the pion daughters of the $f^0(980)$ decay: $p_T >$ 1 GeV and $|y|<$1. Even though the STAR experiment has particle identification capabilities below this, a higher-$p_T$ cut will naturally reduce the background. With these cuts the signal is reduced to 22 nb. The resulting signal and background $\pi^+\pi^-$ invariant mass distributions are shown in Fig.~\ref{fig:f0signalstar}. Even though the $f^0(980)$ cross section into STAR is comparable to the $a^0(980)$ into PHENIX, the $\pi^+\pi^-$ background is larger. An integrated signal-to-background is $4\times10^{-4}$. The inset shows the signal-to-background over the entire mass range.

With these upper estimates of the signal-to-background values in hand, we estimate what they will be in heavy ion collisions. We assume the production of high-$p_T$ particles, $\phi$, $\pi^0$, $\eta$ and $\pi^{\pm}$, will scale as the number of nucleon-nucleon collisions, $N_{coll}$, in the event. Therefore, we would naively estimate that the signal, which will scale as $N_{coll}$, divided by the background, which scales as $N_{coll}^2$, would be reduced by a factor of $N_{coll}$. In mid-central collisions (20-60\%), a Glauber model yields an $N_{coll}$ of 200. This would be the reduction in the signal-to-background with no additional sources of these mesons in heavy ion collisions.

There is an additional increase in the $a^0(980)$ and $f^0(980)$ signal from recombination. We follow the estimation technique in Ref.~\cite{Fries:2003vb} to obtain an order-of-magnitude estimate of the enhanced yield of tetraquarks due to recombiation. In that paper, the yield of mesons from recombination is related to products of the quark and anti-quark Wigner functions. For tetraquark states, the yield from recombination will be related to the products of two quark and two anti-quark Wigner functions. Assuming that the Wigner functions are exponential and independent of the quark or anti-quark internal quantum numbers, the $a^0(980)/\eta^\prime$ will be a ratio of sums over the valence quark flavor, color, and spin. Using the notation of Ref.~\cite{Fries:2003vb} this is
\begin{eqnarray}\label{eq:4qtomesonratio}
\frac{dN_{q\bar{q}q\bar{q}}}{dN_{q\bar{q}}} = \frac{\sum_{a,b,c,d}}{\sum_{a,b}}
\end{eqnarray}
where $a$, $b$, $c$ and $d$ represent the internal quantum numbers of valence quarks in the meson. This ratio is on the order of 10. The estimate assumes that the $\eta^\prime$ mesons will come predominantly from recombination. At 6 GeV, where we would like to apply this ratio, this might not seem to be a good assumption. However, as seen in Fig.~\ref{fig:a0f0v2} and \ref{fig:a0f0v2_minbias}, the tetraquark meson production mechanism will be dominated by recombination up to this point. We have also neglected the reduction of the $\eta^\prime$ production due to energy loss~\cite{Adare:2010dc}. Including this estimate of the increase in yield due to recombination, the signal-to-background will decrease by a factor of 20 compared to $p+p$ collisions.

One could expect that there is additional signal from $\pi^0-\eta$ and $\pi-\pi$ rescattering in the hadronic state, which would enhance the yield of these mesons and would mimic a 4-quark $v_2$ signal. We have checked this contribution using UrQMD \cite{Bleicher:1999xi} which simulates both $a^0(980)$ and $f^0(980)$ resonance production in the hadronic state. We ran UrQMD in minimum bias mode and determined the absolute cross section for each resonance at 6 GeV. For this we used $\sigma_{AuAu}$ = 6.4 b \cite{Miller:2007ri}. For the $a^0(980)$ we determined a mid-rapidity cross section requiring the daughters to decay to photons and require the same kinematic cuts into PHENIX as we described above. We find a cross section of of 0.9 $\mu$b. This should be compared with the $N_{coll}$- and recombination-enhanced cross sections fromx the the $p+p$ production. From above the the $\phi$ and $\eta^{\prime}$ bound the cross section in $p+p$ into the PHENIX between 2-22 nb. In Au+Au from $N_{coll}$ and recombination we expect the production cross section to be 4-44 $\mu$b. The lower bound is a factor of 4 above the hadronic-rescattering cross-section. Similarly, the $f^0(980)$ 6 GeV cross section into STAR using the same kinematic cuts as before was found to be 0.1 $\mu$b. The cross-section bounds on the $f^0(980)$ in $p+p$ collisions is between 14-22 nb. In Au+Au we would expect this cross section would be between 30-44 $\mu$b. Therefore, for $f^0(980)$ production, the hadronic-scattering component is negligible. For the $a^0(980)$, it is a small component.

\section{Discussion}\label{sec:Discussion}
From the previous section it is clear that the measurement of these resonances in heavy ion collisions will be difficult. Suitable background rejection must take place. An example would be placing higher $p_T$ cuts on the daughters in order to reduce the combinatorial background. However, increasing the $\pi^0$ cut to be greater than 2 GeV when considering $a^0(980)$ decays in PHENIX, will only increase the signal-to-background by 30\%. One might wish to search for the radiative $\phi$ decays in heavy ion collisions. Requiring that the resonance combined with a photon be in the $\phi$ mass range will certainly reduce the background significantly. However, the mesons from the radiative decays would not have the flow signature of interest. Because of the small Q of the decay, the $a^0(980)$ or $f^0(980)$ would generally be aligned with the high-$p_T$ $\phi$ and exhibit the 2-quark $v_2$ signature of the $\phi$. Potentially, one could measure inclusive $a^0(980)$ and $f^0(980)$ and compare with those from $\phi$ decay since their $v_2$ signatures would be different. However, this would require a long running time to achieve the necessary statistics for such a measurement.

This idea could be extended to other narrower tetraquark candidates, for example the $X(3872)\rightarrow J/\psi\pi^+\pi^-$. The advantage of measuring this state would be the narrow 3 MeV width~\cite{Nakamura:2010zzi}. The drawbacks for this measurement at RHIC is that the cross section is quite low considering the rate that was observed at Fermilab in inclusive production~\cite{Aaltonen:2009vj,Abazov:2004kp}. Furthermore, because the charm quark is massive compared to light quarks, its $v_2$ is likely not the same as the light quarks~\cite{Greco:2003vf,Zhang:2005ni}. The charm quark $v_2$ has not been measured directly at RHIC, but only indirectly from the electrons from semi-leptonic decays~\cite{Adare:2006nq,Adare:2010de}. In that measurement there is a mixture of bottom and charm sources~\cite{Adare:2009ic,Aggarwal:2010xp}. At increased RHIC luminosity, however, it may be possible to measure this resonance. Such a measurement might be more feasible at the Large Hadron Collider where the rate is higher.

From the empirical evidence for constituent quark scaling, we expect a large $v_2$ for the exotic scalar mesons $a^0(980)$ and $f^0(980)$ in $\sqrt{s_{NN}}$ = 200 GeV Au+Au collisions at RHIC if they are tetraquark states. Their $v_2$ reaches a value of 0.38 in 20-60\% collisions at 6 GeV and is a factor 2.6 higher than the 2-quark $v_2$ at the same $p_T$. The mid-rapidity cross section of these mesons in $p+p$ collisions will be larger than 19 nb for the $f^0(980)$ and 3 nb for the $a^0(980)$, which results from their radiative $\phi$ decays. This is not a negligible cross section at RHIC. Because of the broad mass structure, the signal-to-background is at and below 0.001 over the full 300 MeV mass range of the resonances. The signal-to-background will further be reduced in heavy ion collisions due to the increase in the underlying event multiplicity. Even with the enhanced production due to recombination, the signal-to-background will decrease by a factor of 20. Therefore, without taking steps to increase the signal-to-background, the measurement will be difficult.

\section*{Acknowledgements}
The authors would like to thank Dr.~Rainer Fries for useful discussion. One of us (MW) was supported by Augustana College.

\bibliography{a0paper}

\end{document}